\documentclass[12pt,preprint]{aastex}

\slugcomment{Revised Manuscript; 28 Sep 2005}

\newcommand{\rxte}{{\it RXTE}}
\newcommand{\chandra}{{\it Chandra}}
\newcommand{\sax}{{\it BeppoSAX}}
\newcommand{\xmm}{{\it XMM-Newton}}
\newcommand{\swift}{{\it Swift}}
\newcommand{\rosat}{{\it ROSAT}}
\newcommand{\ergcmsqdsec}{{erg cm$^{-2}$ s$^{-1}$}}
\newcommand{\etal}{et al.}

\begin{document}

\title{The {\it Swift}/BAT High Latitude Survey: First Results}
\author{C. B. Markwardt,\altaffilmark{1,2}
        J.    Tueller,\altaffilmark{2}
        G. K. Skinner,\altaffilmark{3}
        N.    Gehrels,\altaffilmark{2}
        S. D. Barthelmy,\altaffilmark{2}
        R. F. Mushotzky\altaffilmark{4}
        }

\altaffiltext{1}{Department of Astronomy, University of Maryland,
        College Park, MD 20742; craigm@milkyway.gsfc.nasa.gov}
\altaffiltext{2}{Astroparticle Physics Laboratory,
        Mail Code 661, NASA Goddard Space Flight Center, Greenbelt, MD 20771}
\altaffiltext{3}{Centre d'Etude Spatiale des Rayonnements,
        CNRS/UPS, 9 Avenue du Colonel Roche, 31028 Toulouse Cedex 04, France}
\altaffiltext{4}{X-ray Astrophysics Laboratory,
        Mail Code 662, NASA Goddard Space Flight Center, Greenbelt, MD 20771}

\begin{abstract}
We present preliminary results from the first 3 months of the
\swift\ Burst Alert Telescope (BAT) high galactic latitude survey
in the 14--195 keV band. The survey reaches a flux of
$\sim10^{-11}$ \ergcmsqdsec\ and has $\sim 2.7$\arcmin\ (90\%
confidence) positional uncertainties for the faintest sources.
This represents the most sensitive survey to date in this energy
band. These data confirm the conjectures that a high energy
selected AGN sample would have very different properties from
those selected in other bands and represent a `true' sample of the
AGN population.
        We have identified 86\% of the 66
high-latitude sources. 12 are galactic type sources and 44 can be
identified with previously known AGN. All but 5 of the AGN have
archival X-ray spectra, enabling the estimation of line of sight
column densities and other spectral properties. Both of the
$z>0.11$ objects are Blazars. The median redshift of the others
(excluding radio-loud objects) is 0.012. We find that the column
density distribution of these AGN is bimodal with 64\% of the
non-blazar sources having column densities $N_H \geq 10^{22}$
cm$^{-2}$. None of the sources with $\log L_X>$43.5 (c.g.s. units) show high
column densities and very few of the lower $L_X$ sources have low
column densities. Based on these data, we expect the final BAT
catalog to have $>$200 AGN and reach fluxes of less than
$\sim10^{-11}$ \ergcmsqdsec\ over the entire sky.

\end{abstract}


\section{Introduction}

It is now realized that most AGN have high column densities of
absorbing material in our line of sight which significantly change
their observable properties across much of the electromagnetic
spectrum. This material can effectively hide the soft X-ray,
optical and UV signatures of an active nucleus. There are two
spectral bands, the hard X-ray ($E>20$ keV) and the mid-IR (5--50
$\mu$m), where this obscuring material is relatively optically
thin for column densities less than $10^{24}$ cm$^{-2}$. Thus
these bands are optimal for unbiased AGN searches
\citep{treister05}. Recent observations with the Spitzer
observatory are revealing many AGN via their IR emission
\citep{stern05},  but this process is hampered by strong emission
from star formation and the lack of a unique spectral signature to
separate AGN from normal galaxies.

There has been little progress in hard X-ray surveys in the last
25 years due to a lack of instruments with sufficient angular
resolution to identify counterparts in other wavelength bands and
with sufficient sky coverage and sensitivity to produce a large
sample. INTEGRAL observations are predominantly in the galactic
plane and  high latitude coverage is patchy.

We report here preliminary results from the \swift\ BAT high
galactic latitude ($|b|>19\arcdeg$) survey. Although this report
is based on only the first 3 months, we detect 66 sources above
$5.5\sigma$ significance. Only 9 of the sources do not have firm
identifications. Being basically unaffected by obscuration, BAT
determines the intrinsic 14--195 keV luminosity, $L_X$. The survey is already
about ten times more sensitive than the previous large solid angle
survey in this band from HEAO-1 A4 \citep{levine84} and covers the
whole sky at $\sim 1-3\times 10^{-11}$ \ergcmsqdsec.

\section{Observations}

\begin{figure}[h]
\plotone{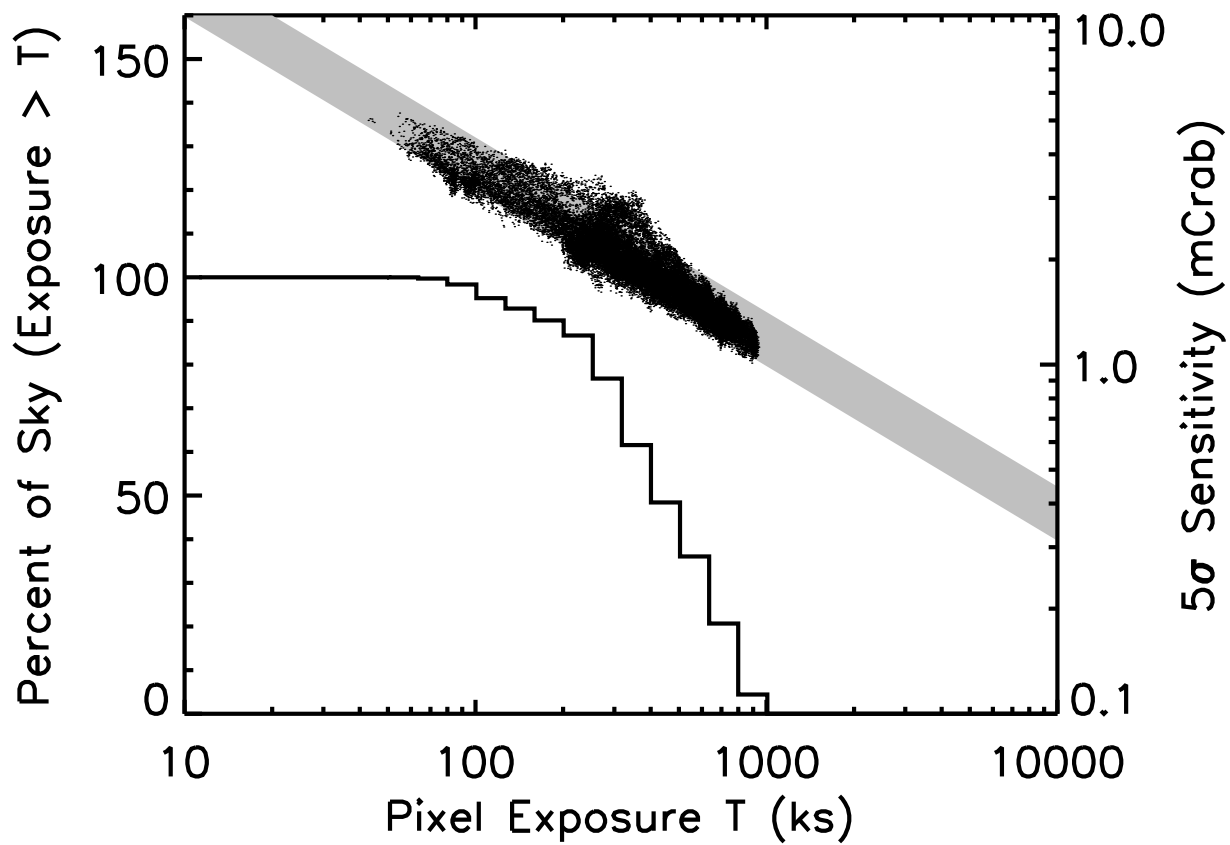}
\figcaption { BAT all-sky sensitivity and exposure statistics. Dots
(top; right scale) represent the $5\sigma$ pixel sensitivity
threshold. The gray band corresponds to the function $(8.5 \pm
1.5) {\rm mCrab} (T/20 {\rm ks})^{-0.5}$, extrapolated to larger
exposures $T$.  The solid line (bottom; left scale) is the
cumulative sky coverage with an exposure of greater than $T$,
expressed as a percentage of the solid angle analyzed
(100\% = $0.67\times 4\pi$).
\label{fig:sens}}
\end{figure}

\begin{figure}
\plotone{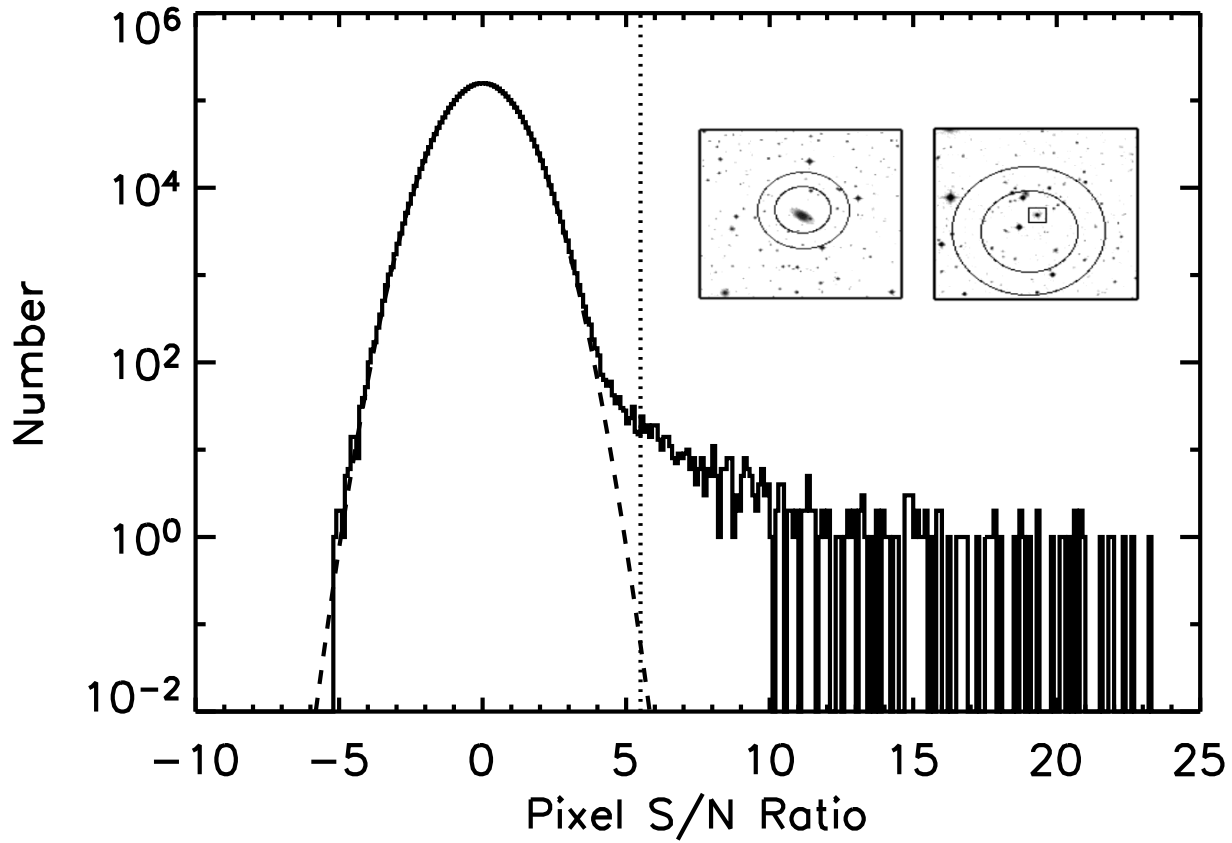}
\figcaption { Distribution of BAT all-sky map pixel significances.
The best-fit gaussian curve (dashed) has a mean and standard
deviation of 0.002 and 1.009, respectively.  The vertical dotted
line is the pixel detection threshold of $5.5\sigma$. Insets:
example BAT-detected sources overlaid on DSS images centered on
MCG $-$5$-$23$-$016 (left) and 3C 390.3 (right), respectively.
The images are 8\arcmin\ on a side (east is left, north is up).
The larger circles represent the BAT 90\% and 99\% confidence
regions for significances of $26\sigma$ (MCG $-$5$-$23$-$016) and
$10\sigma$ (3C 390.3).  The small box identifies 3C 390.3.
\label{fig:resid}}
\end{figure}

The BAT instrument on \swift\ is a very large field of view coded
aperture hard X-ray telescope with a CdZnTe detector array ($\sim$1/8
of the sky is at least partially coded at any one time). While
primarily designed for the detection and rapid dissemination of
gamma-ray burst positions, the BAT is also an effective all-sky hard
X-ray monitor and survey instrument. The BAT focal plane consists of a
0.5 m$^2$ CdZnTe array, divided into 32768 detectors, providing good
sensitivity and energy resolution in the 14--195 keV energy range. BAT
can reach $\sim$65 mCrab in a typical 450 s integration and typically
covers 50--80\% of the sky each day. The effective exposure during the
first 3 months varies over the sky from 200 to 800 ks. Away from the
galactic center region, the sensitivity scales as the square root of
the exposure and so varies by a factor two (Figure
\ref{fig:sens}). The statistical quality of the BAT survey map can be
assessed by comparing positive and negative fluctuations. Figure
\ref{fig:resid} shows the excess in positive fluctuations of
individual pixels. With the reconstruction algorithm used, each source
appears in several correlated pixels; the source detection
significance is approximately the peak value. Since the noise
distribution is symmetric, and there are no pixels $<-5.5\sigma$, we
expect no false detections above our $+5.5\sigma$ threshold. We
estimate that the sensitivity limit is $<$0.5 mCrab for $\sim$50\% of
the sky.  

The survey is based on individual sky images produced using  FFTs
to correlate the data from one pointing (typically a few 1000 s)
with the mask pattern. The FFT is oversampled by a factor of 2 to
prevent loss of sensitivity at bin boundaries. Each image is a
tangent plane projection with a point spread function (PSF) of
22\arcmin\  full-width at half maximum at the image center. A
single image has very non-uniform sensitivity due to partial
coding at the edges (we exclude regions with $<15\%$ partial
coding). Background variations around the orbit and other
systematic effects cause the noise to vary both spatially and
temporally. To achieve a good sensitivity it is necessary to clean
from the data the effects of bright sources and of constant
background non-uniformities. The noise in the resulting cleaned
images is often near the statistical limit. However, some excess
noise remains, so uncertainties used in this analysis are
calculated from the observed r.m.s. noise in the images.  These
images are interpolated and combined to form an all-sky map with
5\arcmin\ sampling.  Pixel weights are based on the local r.m.s.
noise in the component images. The r.m.s. error is then
recalculated and the image is searched for local maxima greater
than $5.5\sigma$. Each maximum is least squares fitted with a
gaussian PSF to derive a position and flux.

The output from two separate analysis pipelines (developed by authors
C. M. and J. T. vs. G. S.), using somewhat different background
correction and image combination algorithms, have been compared and
they agree well.

Figure \ref{fig:resid} shows the BAT confidence contours plotted
on the images of two known sources, illustrating the variation in
position accuracy with source significance. The systematic and
statistical errors are estimated by comparing $\sim$1800 BAT
position measurements with the known positions for $\sim$60 X-ray
sources. The precision improves from $\sim$3.7\arcmin\ (95\%
confidence radius) for sources at $\sim6\sigma$ significance to
$\sim$0.9\arcmin\ at $>20\sigma$ significance. For these early
data, the flux calibration is uncertain by $\sim$30\% as a result
of the spectrally dependent count rate to flux conversion. At
present, the calculation of exposure as a function of sky position
is still approximate. Consideration of the $\log N-\log S$
distribution and of source spectra are thus left to a later paper.

\section{Results}

Our list contains 66 sources at $|b|>19\arcdeg$, of which 12 can
be identified as galactic or SMC/LMC objects, and 45 have clear
identifications with cataloged optical objects, (e.g. the catalog
of \citet{veron03}).  Except for the Coma cluster, all 45 are
known AGN (Table \ref{tab:agn}).  Of the 9 remaining detections,
we have tentative identifications for four cataloged objects which
do not have bright \rosat\ counterparts. Two of the unidentified
detections are \rxte\ slew survey sources (XSS J05054$-$2348 and
XSS J12389$-$1614, which have nearby bright galaxies). Three
sources do not have obvious optical counterparts. All but five of
the cataloged AGN (ESO 297$-$018, NGC 1142, MCG +04$-$22$-$042,
ESO 323$-$077 and Mrk 1498) have published or available X-ray
spectra.

The identified sources are dominated by low luminosity, low
redshift objects. There are three Blazars (3C 273, 4C +71.07, and
Mrk 421) and five radio loud AGN (3C 390.3, Cen A, 3C 120, 4C
+74.26 and Mkn 1498), consistent with the roughly 10\% of all AGN
that are thought to be radio loud. The line of sight column
densities and X-ray fluxes of the 39 objects (3 Blazars and 36
Seyferts) with X-ray spectra have been obtained primarily by ASCA
or \sax\ \citep[e.g.,][]{lutz04}. 
In those cases where Lutz \etal\
do not quote a column density, we have used results from the
Tartarus ASCA data base or from archival \xmm\ or \chandra\ data.
Because the Tartarus data base uses only simple spectral models,
the precision of column densities is only $\pm0.5$ dex. However
this is sufficient to categorize the objects as heavily absorbed
or not.

Assuming a typical power law spectrum ($\Gamma \sim 1.7$), a
conservative absorption column ($\log N_H = 24$), and the BAT limiting
flux, we derive a 2--10 keV limiting flux of $\sim 10^{-12}$
\ergcmsqdsec.  Using the flux distribution of \citet{moretti03}, we
expect $<0.01$ chance sources per BAT error circle, and hence $<1$
misidentification overall.

\begin{figure}
\plotone{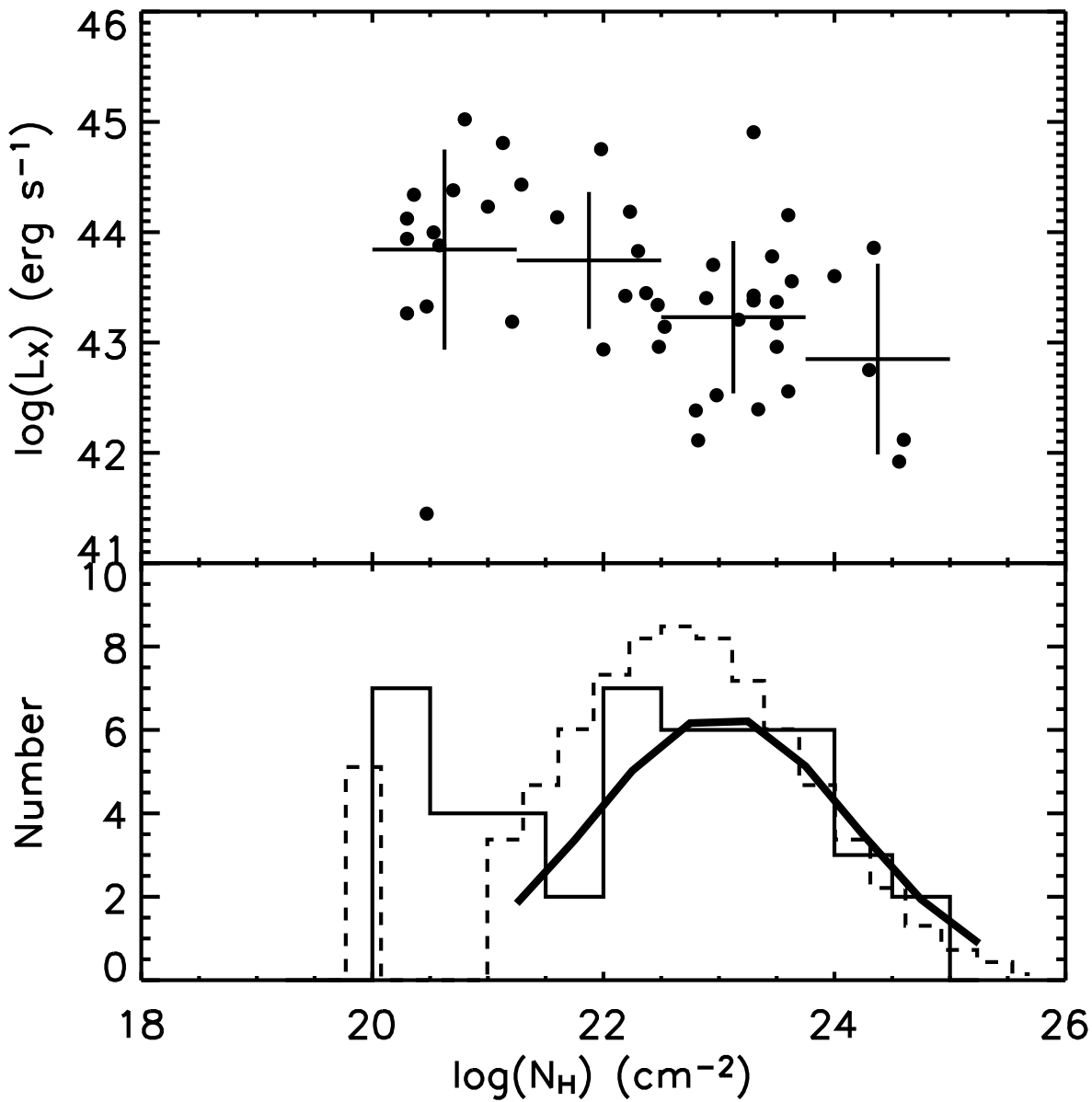}
\figcaption { Correlation between absorption and BAT hard X-ray
luminosity (dots; top).  The crosses show mean and standard
deviation of the luminosity in several $N_H$ bins. The
distribution of absorptions for BAT detected sources is shown in
the bottom panel (solid line). Also shown are the distribution
predicted by \citet{hopkins05a} (dashed), and a gaussian fit to
the bins for $\log N_H > 21$ (solid; centroid = 23.0; $\sigma$ =
1.1).  The peaks at $\log N_H \sim 20-21$ are artificial: for the BAT
sources, the peak is an upper limit to the measureable absorption 
in the 2--10 keV band; and for Hopkins, who collect all sources 
for $\log N_H < 21$ into a single bin.
\label{fig:nh_histo}}
\end{figure}

Excluding the Blazars, the median redshift is $z\sim 0.012$ (the
mean is 0.018), giving a median $\log L_X ~ 43.3$
in the 14--195 keV band and a median column density
of $\log N_H$ = 22.6. The X-ray luminosity against column density
scatter plot for the non-blazars (Figure \ref{fig:nh_histo}) shows
an absence of high column density, highly luminous sources.

A surprising result from our survey is the very high
fraction of identified sources. In HEAO-1 \citep{piccinotti82},
$\sim$1/2 of the 2--10 keV X-ray selected AGN were `new' objects.
In the deep \chandra\ surveys \citep{barger05}, $\sim$1/3 of all
sources did not show a strong optical AGN signature. We thus
anticipated that many of the BAT hard X-ray selected objects would
not have a cataloged AGN counterpart. The high identification rate
of this first BAT sample shows that, contrary to some suggestions,
classical optical techniques have been successful at finding
objects with high line of sight column densities in the low
redshift universe.

Defining an absorbed object as one with  $\log N_H\geq 22.0$
\citep{ueda03}, the ratio of absorbed to unabsorbed objects is
1.8:1 (excluding the Blazars), somewhat less than suggested by the
standard `unified' AGN models. The distribution of column
densities is approximately bimodal (Figure \ref{fig:nh_histo}).
The fact that a ratio of 1:3 was
found by \citet{sazonov05}  for identified sources in the \rxte\
slew survey indicates the strength of the bias towards the
detection of unobscured objects in 2--10 keV low redshift X-ray
surveys.

The distribution of optical classes of the BAT sources is very
different from those in an optical color selected or line selected
survey, with only 20\% of the objects being optically classified
as Seyfert Is, compared with $\sim$40\% in optical surveys at low
redshift. This illustrates the power of a hard X-ray survey to
find all classes of AGN.

Given the limited sensitivity of the BAT survey (all but 15\% of
the sources are brighter than $\sim3\times 10^{-11}$
\ergcmsqdsec), it is surprising that many are not in the HEAO-1 A2
catalog \citep{piccinotti82}. Most of the 18 objects not seen by
HEAO-1 were clearly missed due to obscuration, since 12 of them
have column densities greater than $10^{23}$ cm$^{2}$. Three of
these `missing' objects (Akn 120, 3C 390.3 and 4C +74.26) are
known to be highly variable. Akn 120 is the only narrow line
Seyfert~I in the BAT sample, which is consistent with the known
steep high energy spectrum of these objects \citep{brandt97}.

Comparison of the HEAO-1 2--10 keV fluxes with the more recent
\rxte\ fluxes \citep{revnivtsev04} shows a variation of $\sim$60\%
in flux, with only $\sim$10\% of the objects showing more than a
factor of two variability across the $\sim$20 years between these
two data sets. While the \rxte\ slew survey is much more sensitive
than BAT for unabsorbed objects, at least 8
BAT sources were not detected in the \rxte\ data (Mrk 348, Mrk
1498, NGC 3081, UGC 5037, NGC 1142 and NGC 1365). Of these, 3 have
high column densities, and 3 do not have X-ray spectra.

Most (36/44) of the AGN have cataloged \rosat\ fluxes (taken from
the ROSPSPC catalog when available, otherwise taken from the
\rosat\ all sky survey data base). There is little or no
correlation between the BAT and \rosat\ fluxes. This is not
surprising because the scatter in the \rosat\ rates is dominated
by the effect of different degrees of intrinsic absorption in the
sources, whereas these have relatively little effect in the BAT
energy range. This illustrates the difficulty of constructing a
complete AGN sample from soft X-ray data. Since there is a very
strong relation between optical nuclear and soft X-ray flux
\citep{mushotzky04, barger05}, samples based on optical data are
subject to similar difficulties.

A crucial component for  models of the X-ray background  is the
 distribution of $N_H$ values. Our data determine this distribution, in an unbiased way, for the first time.
 There is  little evidence
for variation of $N_H$ with $L_X$ below $\log L_X=$43.5, but above
this luminosity $<\log N_H$$>$ drops to $\sim20.9$ compared with
22.9 below it. This break point occurs around the characteristic
luminosity of AGN in the \chandra\ surveys of $\log L_X \sim 43.8$
(corrected for the band pass differences; \citet{barger05}) and is
probably related to this feature.  The only object at $\log
L_X > 43.5$ with a high line of sight column density is EXO
055620$-$3820.2. This object \citep{crenshaw01} has complex
absorbing material \citep{quadrelli03} that seems to be associated
with the host galaxy. In the entire BAT AGN sample, including low
latitude objects, there are no extra-galactic objects with $\log
L_X < 42.5$ which are not heavily absorbed. This is not a
selection effect. While such objects clearly exist \citep[e.g., NGC 3998][]{ptak04},
they must be rare. The BAT data establish the
$z=0$ relationship of absorption and luminosity needed to model
the evolution of hard X-ray sources e.g. \citep{mateos05}.

Recently \citep{hopkins05a,hopkins05b} have constructed physical
models which apparently can explain much of the observed evolution
of AGN and the differences between samples in different wavelength
bands. These models only correctly predict the observed absorption
distribution when they include cold material in the line of sight.
The predicted distribution of column densities and the predicted
loose correlation between $L_X$ and $N_H$  are in rough agreement
with the BAT data. However this model completely misses the
observed sharp reduction in high $N_H$ objects at high luminosity.

Comparing the BAT luminosities with 10 or 3.5 $\mu$m IR
luminosities \citep[e.g.,][]{lutz04,gorjian04} shows little or no
correlation. A similar lack of correlation of the ratio of the IR
to BAT flux ratios with X-ray absorption indicates a wide scatter
between observed IR properties and the intrinsic properties of the
AGN.  Previous work \citep{krabbe01} had reported a strong
correlation between the 10 $\mu$m IR fluxes and the hard X-ray
fluxes, na\"ively expected if the near IR is dominated by AGN
light which is relatively unabsorbed. However in the new Spitzer
results \citep{francheschini05} such a correlation is only seen
for type I objects. This lack of correlation may be due to
additional IR radiation from star formation, high optical depths
even at 10 $\mu$m, or  additional scatter introduced by
reprocessing of the nuclear radiation in the IR. The absence of a
correlation makes the analysis using the unified models to connect
the IR and X-ray data suspect \citep{treister05}. Similarly there
is little or no correlation between BAT and [OIII] luminosities.

The BAT sample allows a true measure of the nature of the low $z$
hosts of active galaxies. Of the 24 objects in the sample with ``T''
types as categorized in the RC3 \citep{devacouleurs91},
9 have ``T'' types $<$0, indicating a spheroidal
host fraction of $\sim$40\%, similar to that seen in the \chandra\
surveys and the SDSS data \citep{kauffmann05,grogin05}. This is
much larger than the fraction in classical optical surveys.
However, none of the objects is  classified as a giant elliptical
galaxy, which is rather different than the \chandra\ surveys,
probably due to the low redshifts probed by the BAT sample.

It is interesting to note that all the broad line radio galaxies
detected by HEAO-1 \citep{marshall79} (3C 111.0, 3C 120, 3C 382,
3C 390.3) are detected by the BAT, indicating that radio galaxies
may have systematically harder X-ray spectra in the BAT band than
Seyferts.

\section{Conclusions}
We have shown the usefulness of a sensitive, large solid angle
hard X-ray survey in defining the nature of the AGN population. As
predicted by models of the X-ray background, the dominant source
population are absorbed AGN and are very different from an
optically or soft X-ray derived AGN population. We  derive the
true distribution of absorbing column density with X-ray
luminosity and find that while virtually all sources with $\log
L_X<$43.5 are absorbed, those with higher luminosities are mostly
unabsorbed. This luminosity corresponds to the break in the X-ray
luminosity function and is a strong clue to the nature of the
absorbing material and the origin of the feature in the luminosity
function. We  find little correlation between other measures of
the luminosity of absorbed AGN (e.g. [OIII], 3.5 or 10 $\mu$m
luminosity) and the hard X-ray luminosity, suggesting that
previous techniques may have previously unsuspected biases in
finding and measuring absorbed AGN. The high identification
fraction ($\sim$86\%) is somewhat unexpected given our previous
knowledge of these sources and bodes well for  the much larger
samples that the BAT will obtain over the next two years.

Further work on the present sample will include determination of
the $\log N-\log S$ of the hard X-ray sample,the luminosity
function, spectral and temporal analysis of the BAT selected AGN
directed towards a detailed understanding of the relationship of
the hard X-ray to other wavelength bands. Preliminary estimates
indicate that an eventual 3(+) year BAT high latitude catalog will
have more than 200 AGN, which will allow extensive correlation
analyses.

\acknowledgments We thank the entire BAT and {\it Swift\/} teams for their
extensive efforts which have made this work possible.

\clearpage

\begin{deluxetable}{lrrrlrrrrrcl}
\tablenum{1}
\tabletypesize{\tiny}
\tablecaption{BAT Detections of AGN (On-line only table)\label{tab:agn}}
\tablehead{
\colhead{SWIFT Name} & 
\multicolumn{2}{c}{R.A.\hfil(J2000)\hfil Decl.} & 
\colhead{Err.\tablenotemark{a}} & 
\colhead{Identification} & 
\colhead{Offset} & 
\colhead{$z$} & 
\colhead{$\log N_H$} & 
\colhead{Ref.} &
\colhead{$\log L_X$\tablenotemark{b,c,d}} &
\colhead{RL\tablenotemark{e}} & 
\colhead{Type\tablenotemark{f}} \\
\colhead{} & \colhead{\arcdeg} & \colhead{\arcdeg} & \colhead{\arcmin} &
\colhead{} & \colhead{\arcmin} & \colhead{} & \colhead{cm$^{-2}$} &
\colhead{} & \colhead{erg s$^{-1}$ cm$^{-2}$} & \colhead{} & \colhead{}}
\startdata
 J0048.8$+$3157&   12.200&   31.963& 4.8&             Mrk 348& 0.4& 0.0150&      22.9&         1& 43.7&  N& Sy\\
 J0123.7$-$3502&   20.942&  -35.040& 4.4&            NGC 526A& 2.3& 0.0191&      22.3&         2& 43.8&  N& Sy\\
 J0123.9$-$5848&   20.978&  -58.811& 4.9&           Fairall 9& 1.2& 0.0470&      20.4&         2& 44.3&  N& Sy\\
 J0138.9$-$4005&   24.730&  -40.094& 5.3&       ESO 297$-$018& 6.0& 0.0252&   \nodata&   \nodata& 43.4&  N& Sy\\
 J0238.2$-$5216&   39.562&  -52.275& 5.2&       ESO 198$-$024& 5.0& 0.0455&      21.0&         1& 44.2&  N& Sy\\
 J0255.0$-$0009&   43.765&   -0.158& 4.7&            NGC 1142& 2.6& 0.0288&   \nodata&   \nodata& 44.3&  N& Sy\\
 J0333.5$-$3608&   53.377&  -36.143& 3.2&            NGC 1365& 1.2& 0.0055&      23.6&         2& 42.6&  N& Sy\\
 J0433.1$+$0521&   68.288&    5.362& 3.7&              3C 120& 0.7& 0.0330&      21.2&         1& 44.4&  Y& Sy\\
 J0451.8$-$0347&   72.964&   -3.799& 5.1&MCG $-$01$-$13$-$025& 2.5& 0.0159&      20.5&         3& 43.3&  N& Sy\\
 J0516.3$-$0008&   79.078&   -0.137& 4.6&             Ark 120& 2.0& 0.0323&      20.3&         2& 43.9&  N& Sy\\
 J0557.8$-$3822&   89.466&  -38.371& 3.8& EXO 055620$-$3820.2& 3.0& 0.0339&      22.2&         1& 44.2&  N& Sy\\
 J0615.9$+$7100&   93.996&   71.000& 3.3&               Mrk 3& 2.9& 0.0135&      24.0&         2& 43.6&  N& Sy\\
 J0841.7$+$7052&  130.437&   70.875& 4.5&         4C $+$71.07& 2.1& 2.1720&      21.0&         1& 48.3&  Y& Bl\\
 J0924.1$+$2256&  141.038&   22.945& 4.7&MCG $+$04$-$22$-$042& 6.4& 0.0323&   \nodata&   \nodata& 43.9&  N& Sy\\
 J0924.8$+$5216&  141.220&   52.280& 4.2&             Mrk 110& 3.1& 0.0353&      20.6&         1& 43.9&  N& Sy\\
 J0945.6$-$1420&  146.411&  -14.343& 4.1&            NGC 2992& 1.3& 0.0077&      22.0&         4& 42.9&  N& Sy\\
 J0959.4$-$2250&  149.874&  -22.848& 4.4&            NGC 3081& 1.3& 0.0080&      23.5&         5& 43.0&  N& Sy\\
 J1023.5$+$1952&  155.875&   19.870& 3.2&            NGC 3227& 0.3& 0.0039&      22.8&         6& 42.4&  N& Sy\\
 J1031.6$-$3450&  157.924&  -34.836& 4.1&            NGC 3281& 2.4& 0.0107&      24.3&         7& 42.8&  N& Sy\\
 J1104.5$+$3813&  166.129&   38.218& 3.2&             Mrk 421& 0.9& 0.0300&      20.3&         8& 44.1&  Y& Bl\\
 J1106.5$+$7235&  166.631&   72.599& 2.9&            NGC 3516& 2.2& 0.0088&      21.2&         1& 43.2&  N& Sy\\
 J1138.9$-$3744&  174.743&  -37.740& 2.5&            NGC 3783& 0.7& 0.0097&      22.5&         2& 43.3&  N& Sy\\
 J1209.6$+$4343&  182.418&   43.730& 5.0&            NGC 4138& 3.3& 0.0030&      22.8&         3& 42.1&  N& Sy\\
 J1210.5$+$3924&  182.627&   39.412& 1.1&            NGC 4151& 0.5& 0.0033&      22.5&         2& 43.0&  N& Sy\\
 J1225.8$+$1239&  186.452&   12.664& 1.9&            NGC 4388& 0.4& 0.0084&      23.6&         2& 43.6&  N& Sy\\
 J1229.1$+$0203&  187.283&    2.052& 2.3&              3C 273& 0.3& 0.1583&      20.5&         1& 46.1&  Y& Bl\\
 J1235.6$-$3953&  188.909&  -39.894& 1.9&            NGC 4507& 0.9& 0.0118&      23.5&         2& 43.8&  N& Sy\\
 J1239.6$-$0518&  189.900&   -5.304& 4.0&            NGC 4593& 2.6& 0.0090&      20.3&         2& 43.3&  N& Sy\\
 J1306.8$-$4023&  196.706&  -40.386& 4.9&       ESO 323$-$077& 4.7& 0.0150&   \nodata&   \nodata& 43.4&  N& Sy\\
 J1322.5$-$1645&  200.630&  -16.759& 5.0&MCG $-$03$-$34$-$064& 2.5& 0.0165&      23.5&         5& 43.4&  N& Sy\\
 J1325.4$-$4300&  201.362&  -43.014& 1.0&               Cen A& 0.4& 0.0018&      22.7&         1& 42.6&  Y& Sy\\
 J1349.3$-$3018&  207.335&  -30.310& 2.0&            IC 4329A& 0.2& 0.0161&      21.6&         1& 44.1&  N& Sy\\
 J1353.1$+$6916&  208.295&   69.269& 5.1&             Mrk 279& 2.5& 0.0305&      20.5&         1& 44.0&  N& Sy\\
 J1413.2$-$0312&  213.316&   -3.207& 2.3&            NGC 5506& 0.3& 0.0062&      22.5&         2& 43.1&  N& Sy\\
 J1442.4$-$1713&  220.602&  -17.233& 4.5&            NGC 5728& 1.2& 0.0093&      23.5&         9& 43.2&  N& Sy\\
 J1628.0$+$5144&  247.017&   51.747& 4.5&            Mrk 1498& 1.7& 0.0547&   \nodata&   \nodata& 44.5&  Y& Sy\\
 J1838.2$-$6526&  279.553&  -65.438& 3.8&       ESO 103$-$035& 1.0& 0.0133&      23.2&         1& 43.2&  N& Sy\\
 J1842.0$+$7945&  280.514&   79.754& 3.1&            3C 390.3& 1.1& 0.0561&      21.0&         1& 44.8&  Y& Sy\\
 J2043.0$+$7504&  310.763&   75.073& 4.7&         4C $+$74.26& 4.0& 0.1040&      21.6&         3& 45.3&  Y& Sy\\
 J2044.3$-$1045&  311.076&  -10.765& 4.9&             Mrk 509& 3.3& 0.0344&      20.7&         1& 44.4&  N& Sy\\
 J2052.1$-$5703&  313.037&  -57.056& 4.3&             IC 5063& 1.2& 0.0113&      23.3&         1& 43.4&  N& Sy\\
 J2201.8$-$3152&  330.461&  -31.882& 3.7&            NGC 7172& 2.5& 0.0087&      22.9&         1& 43.4&  N& Sy\\
 J2254.0$-$1732&  343.522&  -17.541& 4.8&       MR 2251$-$178& 2.5& 0.0640&      20.8&         1& 45.0&  N& Sy\\
 J2318.4$-$4223&  349.609&  -42.391& 3.6&            NGC 7582& 1.3& 0.0053&      23.0&         1& 42.5&  N& Sy\\
\enddata
\tablecomments{$a$  Error radius (99\% conf.);
               $b$  BAT Luminosity, 14--195 keV;
               $c$  Estimated systematic error: +20\%, $-$50\%;
               $d$  Assumes a {\it WMAP\/} cosmology \citep[$H_{\rm 0} = 71 {\rm km \; s^{-1} \;Mpc^{-1}},\Omega_{M} = 0.27, \Omega_{\Lambda} = 0.73$;][]{spergel03}
               $e$  Radio Loud?;
               $f$  Sy=Seyfert; Bl=Blazar}
\tablerefs{
   (1) \citealt{Tartarus};
   (2) \citealt{lutz04};
   (3) This work ({\it XMM-Newton});
   (4) \citealt{gilli00};
   (5) This work ({\it ASCA});
   (6) \citealt{gondoin03};
   (7) \citealt{vignali02};
   (8) \citealt{perlman05};
   (9) This work ({\it Chandra})
}
\end{deluxetable}


\clearpage

\clearpage

\clearpage

\end{document}